# Neural-networks-based Photon-Counting Data Correction: Pulse Pileup Effect

Ruibin Feng, David Rundle, and Ge Wang, *Fellow, IEEE*

*Abstract*—Compared with the start-of-art energy integration detectors (EIDs), photon-counting detectors (PCDs) with energy discrimination capabilities have demonstrated great potentials in various applications of medical x-ray radiography and computed tomography (CT). However, the advantages of PCDs may be compromised by various hardware-related degradation factors. For example, at high count rate, quasi-coincident photons may be piled up and result in not only lost counts but also distorted spectrums, which is called the pulse pileup effects. Considering the relative slow detection speed of current PCDs and high x-ray flux for clinical imaging, it is vital to develop an effective approach to compensate or correct for the pileup effect. The aim of this paper was to develop a model-free, end-to-end, and data-driven approach in the neural network framework. We first introduce the trigger threshold concept to regularize the correction process. Then, we design a fully-connected cascade artificial neural network and train it with measurements and true counts. After training, the network is used to recover raw photon-counting data for higher fidelity. To evaluate the proposed approach, Monte Carlo simulations are conducted. We simulate four types of PCDs combining the unipolar or bipolar pulse shape with the nonparalyzable or paralyzable detection mode. The simulation results demonstrate that the trained neural network decreases the relative error of count averages of raw data by an order of magnitude while maintaining a reasonable variance over a wide range of attenuation paths.

## I. Introduction

X-RAY imaging especially computed tomography (CT) has played a significant role in modern medicine. Almost exclusively, the x-ray detection technology in all current x-ray imagers is based on energy-integrating detectors (EIDs), which integrates electrical signals generated from interactions between a polychromatic x-ray beam and detection materials over the entire x-ray spectrum. Consequently, all energy-dependent information is lost, and the generated linear attenuation coefficients are not tissue-type specific. Additionally, the signal-to-noise ratio of the detected signal is relatively low limited by the dark current (or known as the electric noise) and Swank noise [1]. Finally, lower energy photons, which carry more contrast information, receive lower weights due to beam hardening, which results in poor-contrast CT images.

Recently, photon-counting detectors (PCDs) have been developed for medical x-ray imaging [2]–[8]. In contrast to EIDs,

R. Feng and G. Wang are with Department of Biomedical Engineering, Rensselaer Polytechnic Institute, Troy, NY, 12180 USA (e-mail: fengr@rpi.edu; wangg6@rpi.edu).

D. Rundle is with JairiNovus Technologies Ltd., Butler, PA, 16001 USA (e-mail: david.rundle@jairinovus.com).

PCDs recognize photons both individually and spectrally with multiple energy windows. With the energy discrimination capabilities, PCDs show great potentials to address the issues with EIDs as mentioned above. First of all, PCDs can reveal the elemental composition of materials and distinguish more than one contrast medium simultaneously, especially useful for K-edge and fluorescence imaging. For example, novel contrast agents such as gold nanoparticles (GNPs) [9], [10] become attractive with PCDs. Second, PCDs have an inherently higher signal-to-noise ratio (SNR) since the dark current and Swank noise are suppressed and do not affect the output signal in any energy window. Based on the simulation study in [11], SNRs of PCDs are improved by up to 90% relative to that of EIDs. Last but not least, the weights of x-ray photons are not biased in PCDs. It means the beam-hardening artifacts can be naturally addressed, and contrast-enhanced images are produced as a low-hanging fruit.

It may appear that PCDs can address almost all of the EID-related problems and produce perfect data. Unfortunately, there exist various degradation factors, such as pulse pileup and charge sharing effects, which distort PCD data to such a degree that EIDs may work better under certain conditions. Taking pulse pileup effect as an example, many of the benefits of PCDs disappear when the input count rate exceeds 20% of the maximum characteristic count rate for contrast detection tasks [12]. Comparing with the typical operational count rates for clinical x-ray CT detectors, which is up to 100 - 1000 Mcps/mm$^2$ (million counts per second per mm$^2$), the maximum count rates of most PCDs are currently still lower than 10 Mcps/mm$^2$ according to the survey [13]. Although recent investigations indicate that a high count rate is not necessary with iterative reconstruction algorithms, the current prototypes of PCDs are still too slow to avoid the pulse pileup effect. Hence, to use PCDs for medical applications, it is critical to develop a scheme to correct or compensate for degradation factors such as the pulse pileup effect.

Over past years, many studies were performed from hardware and/or software perspective. For example, the PCD Medipix3RX [14], [15] applied an anti-charge sharing scheme working through charge summing circuits which connect adjacent $2 \times 2$ pixels. After detecting coincident incidents, it reconstructs the counts by exclusively assigning all the charge to the pixel with the largest detected energy. However, this scheme increases the effective pixel size and the chance of count loss. Other challenges also exist, such as the reduced count rate and addition heat from the complicated circuits. For the pileup problem, hardware-based pileup rejectors were developed to reduce or eliminate the spectral distortion [16],

[17]. A drawback with pileup rejection is a significantly decreased precision of measurement due to ignoring counts involved in the pileup effect. On the other hand, photon-counting data processing algorithms were proposed to handle pulse pileup [18], [19] and spectral distortion [20], [21]. The problem is that most of these methods are model-dependent. To be specific, the precise model of the physical detection process is needed, involving multiple factors such as pulse shape, pileup order, and response mechanism. These are difficult to acquire, and when these are formulated, they are often inaccurate. Besides, these schemes are associated with specific tasks such as material decomposition or attenuation estimation. Instead of recovering the true measurement, they integrate the physical model into the problem-solving process. These two limitations significantly reduce the flexibility and generality.

To improve the existing correction/compensation approaches, here we propose a neural-network-based approach to handle the measurements compromised by the pulse pileup effect. First, we introduce the concept of trigger thresholds to provide information about the pileup effect and regularize the correction process. Then, we design a fully-connected cascade neural network and train it with PCD data with/without PPE. Thanks to the data-driven and self-learning properties of the neural network, the detector model is no longer required. After proper training, the raw data can be directly corrected from distorted counterparts and applied for image reconstruction and analysis. Furthermore, the neural network model is fairly simple, and the data could be processed in real-time. To validate our approach, Monte Carlo simulations are conducted to generate training data for four different detector models. The neural-network-based correction reduced the mean absolute percentage error of raw data by an order of magnitude while maintaining reasonable variance over a wide range of attenuation paths.

The remaining parts of this paper are organized as follows. Section II provides the background of PCDs and pulse pileup effect (PPE). Section III describes trigger thresholds and fully-connected cascade architecture. Section IV presents the Monte Carlo simulation settings. Section V summarizes simulation results. Finally, Section VI discusses relevant issues and the future work.

## II. BACKGROUND

Most of the photon-counting detectors with energy discrimination capability use compound semiconductors to effectively absorb x-ray photons. The incident photons are directly converted into electron and hole charge cloud pairs. A negative bias voltage is applied to the incident side of the sensor, creates an electric field and forces these charge clouds to pixelated anodes. Then, with application-specific integrated circuits (ASICs), the signals are amplified and shaped in the form of pulses, whose heights are compared with the energy threshold and counted if the pulse height exceeds the threshold.

### A. Pulse Shape Model

Different pulse shapers can be built into the ASIXs. Unipolar and bipolar pulses are two common choices. A typical unipolar pulse shaper will generate positive pulses with height proportional to the photon energy and a fixed temporal extent. On the other hand, a bipolar pulse contains a positive main part like the unipolar pulse and a negative tail. The normalized unipolar and bipolar pulses are illustrated in FIG. 1. There are some important parameters to model the shape. One is the peaking time denoted as $t_1$ in FIG. 1, which is from incident time to the time where the pulse researches its maximum. Another is pulse width ($t_2$), which is the period from incident time to the time where the pulse height back to the baseline. For a bipolar pulse, we also need to know when the negative tail rise back to baseline ($t_3$) and the height of the tail ($b$). It should be mentioned that there is a minimum amount time required to separate two consequences pulses instead of recording them as one distorted pulse. This time is normally referred as deadtime $\tau$ (or characteristic count rate $1/\tau$ if we are talking its inverse. In most cases, the deadtime is roughly equal to the pulse width.

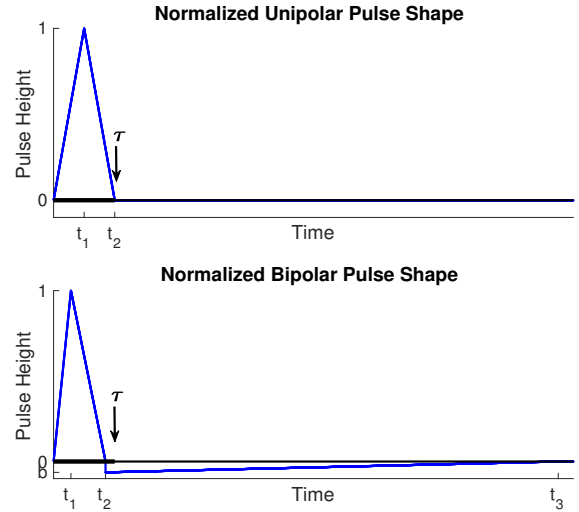

Fig. 1. Normalized unipolar and bipolar pulse shape. The pulse height scales with photon energy while the pulse time is the same. The dead time of the detector is denoted as $\tau$.

### B. Detection Mechanism

When the detector is in the active state (which means that the detector is ready to detect new incident photons), the first arriving photon will set the detector to inactive state which will last for a duration of deadtime. Depending on how to reset detector from the inactive state to the active state, the PCDs can be loosely categorized as nonparalyzable and paralyzable ones, although most of the detectors behave somewhere between [22]. For nonparalyzable detectors, every inactive state will last at most the deadtime, and then the detectors will return to active state after the deadtime or when the signal height drops back to the baseline. This detection mechanism is shown in FIG. 2 (b). On the other hand, the state of a paralyzable detector, illustrated in FIG. 2 (c), returns to the active state only when the pulse drops back to the baseline. It means the period of the inactive state is not fixed, and the detector returns to the active state in the deadtime after an incident photon if and only if there is no additional incident photon during the deadtime. In both detector models, all photons incident on the inactive detector will potentially contribute to the detected pulse signal. This phenomenon is called pulse pileup effect, which will be elaborated in next section.

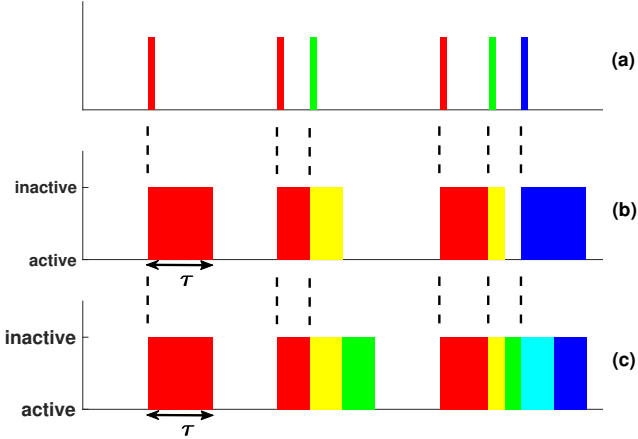

Fig. 2. Example of different detection mechanisms. (a) Incident photons, (b) The active and inactive states of a nonparalyzable detector, and (c) The active and inactive states of a paralyzable detector. Note that we use red, green and blue colors to distinguish photons, and the mixed colors to indicate a combined contribution to the detected signals.

### C. Pulse Pileup Effect

Due to the existence of deadtime and inactive state described in the previous section, the pulse generated by quasi-coincident photons may be piled up and recorded as one pulse with a wrong registered energy. This effect will result in lost counts and a distorted energy spectrum. FIG. 3 illustrates the pulse pileup effect, where the pulses from the 3rd and 4th photons are overlapped and merged as a single pulse. It is clear the 4th photon is not registered. Furthermore, for the 5th and 6th photons, we have not only count loss but also distorted recorded energy. The phenomenon that pulses overlapping for coincident photons is recorded as a single count at a higher energy than the energies of the original pulses is called the peak pulse pileup [16], [17]. If the bipolar pulse shaper is used, a pulse peak may overlap the tail of a preceding pulse and result in a smaller recorded energy. This effect is called tail pulse pileup [16]. It should be noticed that tail pulse pileup could also cause lost counts (e.g. 7th photon in FIG. 3) and energy distortion (seeing 8th photon as an example).

The recorded spectrum can be heavily distorted by the pulse pileup effect, especially when the count rate is much higher than characteristic count rate. The recorded count rate $a_r$ can be expressed as a function of incident count rate $a$ and deadtime $\tau$, which is shown as below:

$$a_r = a \times \begin{cases} 1/(1+a\tau) & \text{for a nonparalyzable detector,} \\ \exp(-a\tau) & \text{for a paralyzable detector.} \end{cases} \quad (1)$$

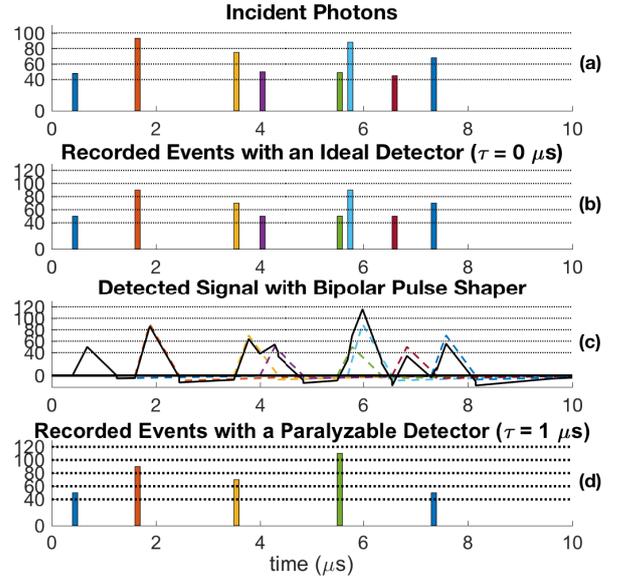

Fig. 3. Example of Pulse Pileup Effect. (a) Incident photons, (b) recorded events using an ideal detector ($\tau = 0\mu s$) with thresholds equal to [40, 60, 80, 100, 120], (c) the detected signal with a bipolar pulse shaper, and the pulses from individual photons are illustrated with dashed lines and the final signal with in solid line, and (d) the recorded events of the signal in (c) with a paralyzable detector ($\tau = 1\mu s$).

The relationship between $a_r$ and $a\tau$ is illustrated in FIG. 4. It can be seen that if the incident count rate is more than 20% of characteristic count rate, $a_r$ will be heavily deviated from $a$. The deviation of the number of recorded counts from the number of real counts can be measured as the deadtime loss ratio (DLR):

$$\text{DLR} = \begin{cases} 1 - 1/(1+a\tau) & \text{for a nonparalyzable detector,} \\ 1 - \exp(-a\tau) & \text{for a paralyzable detector.} \end{cases} \quad (2)$$

As with lost counts, the distortion of the recorded spectrum also increases with DLR. FIG. 5 illustrated the difference between incident spectrum and recorded spectrum with DLR being 1.0061% and 9.8332% respectively.

Based on the above analysis, the direct usage of PCD measurements in applications such as attenuation estimation and material decomposition without compensation or correction may be problematic. In next section, we propose a neural network based correction approach to address the PPE problem.

## III. NEURAL NETWORKS BASED CORRECTION METHOD

### A. Trigger Thresholds

Generally speaking, directly recovery from raw data may not be well-posed problem, since the input and the output are of the same dimensionality. Next, we will introduce the concept of trigger thresholds to tackle this problem. To regularize the data correction process, here we introduce the concept of trigger thresholds. Let us look at FIG. 5 as an example. Without the pulse pileup effect, there should be no counts whose energy is beyond 120 keV for 120 kVp incident x-ray spectrum. However, when DLR is equal to 1%, we start observing "hyper-energy" measurements, such as in energy bin [125 keV, 126 keV]. And when DLR increases to 10%, the mean counts in the same energy bin is equal to 19.17. More information about the increment of the number of counts above the 120 keV threshold is shown in FIG. 6. Note that the ratio is defined between the number of higher-energy counts and the number of recorded counts. Clearly, the measurements from energy thresholds larger than the maximum energy of incident photons (120 keV in our case) provide hints about whether the pulse pileup effect exists and how severe it is. In [19], this idea was applied by introducing the trigger threshold whose energy is larger than x-ray tube energy, while the thresholds within the incident spectrum are called spectral thresholds. Then, PCD data can be corrected by simply adding weighted counts over the trigger threshold ($N_t$) to the data corresponding spectral thresholds ($N_s$). However, this method only takes one trigger threshold into account, and does not provide a scheme to optimize the weight. In our study, we will use both spectral thresholds and more than one trigger thresholds to generate data as the input to the neural network for data correction, as described in the following subsection.

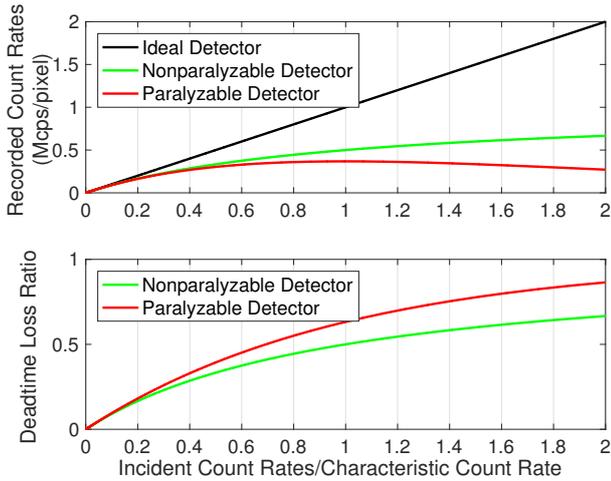

Fig. 4. Recorded count rate and deadtime loss with respect to the ratio between incident count rate and characteristic count rate. The deadtime of nonparalyzable and paralyzable detectors is $1\mu s$ ($\tau = 1\mu s$).

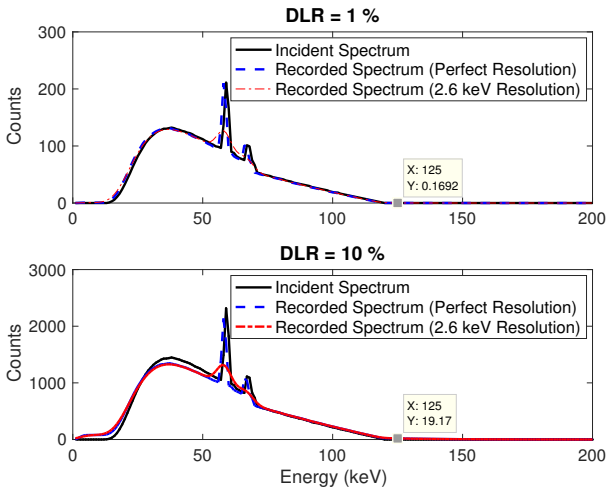

Fig. 5. Incident and recorded spectra with no attenuation and DLR equal to 1% and 10% respectively. A unipolar pulse shaper and a nonparalyzable mode are assumed for an x-ray tube operated at 120 kVp. Data were generated as the mean of 10,000 trails via Monte Carlo simulation described in Section IV by assuming the 120 kVp incident x-ray spectrum, except incident count rate is set as $3.37 \times 10^5$ and $2.7 \times 10^6$ cps/mm$^2$ yielding the desired DLR respectively.

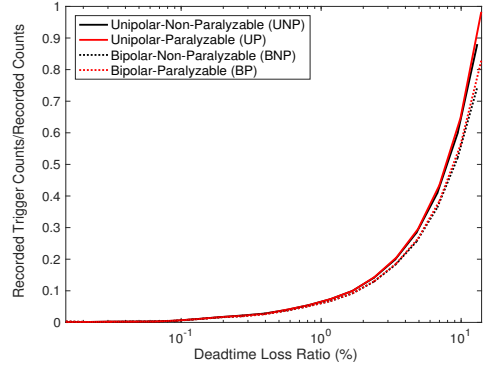

Fig. 6. Ratio between the number of counts above 120 keV threshold and the total number of recorded incidents for various combinations of pulse shapes and detection mode with no attenuation. Data were generated as the mean of 10,000 trails via Monte Carlo simulation described in Section IV assuming the 120 kVp incident x-ray spectrum, except incident count rate is set in the range $[5 \times 10^3, 5 \times 10^6]$ cps/mm$^2$ yielding DLR as $[0.015\%, 13.04\%]$ and $[0.015\%, 13.93\%]$ for nonparalyzable and paralyzable respectively.

### B. Fully-connected Cascade Artificial Neural Network

With the raid growth of computational power and dataset size, machine learning, especially deep neural networks, recently attracted enormous attentions. The main advantage of the neural network lies in its ability to learn from data without modeling the problem especially, making it extremely suitable for correction of PCD data to address the pulse pileup effect.

Since the correction process can be defined as a multivariate regression problem, a natural neural network architecture is the multilayer perceptron (MLP) model. However, training an MLP of several hidden layers could be difficult due to the diminishing gradient problem [23]. In order to address this problem, a fully connected cascade (FCC) architecture, also called the cascade-correlation learning architecture, was proposed [24]. The main difference between FCC and MLP is that any hidden layer is connected to its all preceding layers including the input layer. In this way, connections are made available between different layers, and any layer could directly access the gradients from the loss function. This architecture was successfully applied to solve attention deficit hyperactivity disorder classification problem [25], and extended to build a special type of convolution neural networks, which is referred as DenseNet [26]. The direct connection idea was also implemented in ResNet [27] and Highway Networks [28], which have achieved significant performance gains in many benchmark tasks. Recently, the similar idea was also implemented in the encoder-decoder network for low-dose CT image denoising [29].

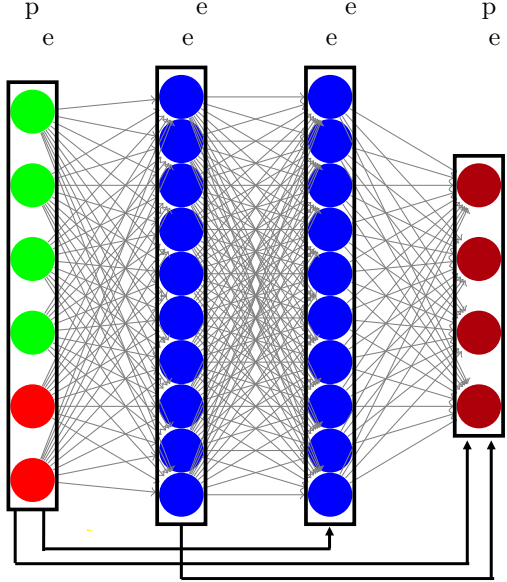

Fig. 7. Fully-connected cascade artificial neural network with 2 hidden layers. The input consists of 4 spectral measurements (green nodes) and 2 triggered measurements (red nodes). The black arrows denote the direct connections from input layer and hidden layer 1 to their subsequent layers.

In this pilot study, we focus on the FCC architecture for data correction. Besides the aforementioned advantage, this architecture facilitates the utilization of data correlation among the input channels and between the input and output data. Specifically, if no PPE is involved, the input without trigger measurements will be identical to output. Hence, FCC is preferable, given the direct connection from the input layer to the output layer. A two-hidden-layer FCC is illustrated in Fig. 7. In the following section, we will detail the Monte Carlo simulations design, which is used to generate training and test datasets.

## IV. MONTE CARLO SIMULATION

We will study representative detector models corresponding to four combinations of different pulse shapes (from unipolar and bipolar pulse shapers respectively) and detection mechanisms (in nonparalyzable and paralyzable modes respectively). These models are referred as Unipolar-Non-Paralyzable (UNP), Unipolar-Paralyzable (UP), Bipolar-Non-Paralyzable (BNP) and Bipolar-Paralyzable (BP) respectively. TABLE I summarizes the parameters used in the simulation. We only consider the single pixel irradiation. The details of each simulation step are given as follows.

### A. Step 1: Incident x-ray spectrum

The 120 kVp incident x-ray spectrum was simulated using the Spektr [30] implementation of TASMICS [31]. The filtra-tion of 2.0-mm aluminum was added. The spectrum ($S_0(E)$) was described as the probability mass function (PMF) with discrete energy windows with a 1-keV-width. It was normalized by the sum of all the component values, as illustrated in FIG. 8.

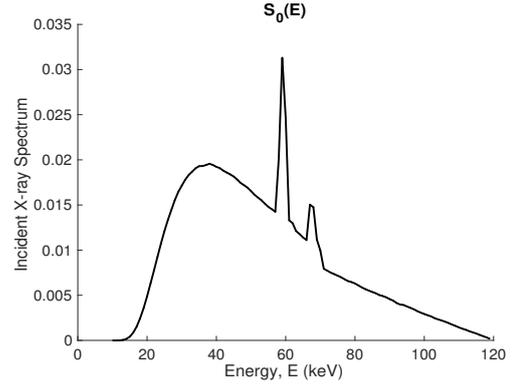

Fig. 8. Incident X-ray spectrum $S_0(E)$.

### B. Step 2: Attenuation

Based on the basic principle of material decomposition, the linear attenuation coefficient $\mu_a(\mathbf{r}, E)$ at position $\mathbf{r}$ and energy $E$ can be approximated by a linear combination of three energy-dependent basis functions: Photoelectric effect $\phi_p(E)$, Compton scattering $\phi_c(E)$, and a third basis function for K-edge imaging $\phi_K(E)$ [3], [32]:

$$\mu_a(\mathbf{r}, E) = c_p(\mathbf{r})\phi_p(E) + c_c(\mathbf{r})\phi_c(E) + c_K(\mathbf{r})\phi_K(E), \quad (3)$$

where $c_p(\mathbf{r})$, $c_c(\mathbf{r})$ and $c_K(\mathbf{r})$ are position dependent basis coefficients of $\phi_p(E)$, $\phi_c(E)$ and $\phi_K(E)$, respectively. Then the line-integrals can be calculated as:

$$\begin{aligned}
\int \mu_a(\mathbf{r}, E) d\mathbf{r} &= \int c_p(\mathbf{r}) d\mathbf{r} \phi_p(E) + \int c_c(\mathbf{r}) d\mathbf{r} \phi_c(E) \\
&\quad + \int c_K(\mathbf{r}) d\mathbf{r} \phi_K(E) \\
&= v_p \phi_p(E) + v_c \phi_c(E) + v_K \phi_K(E) \\
&= \Phi(E)\mathbf{v}, \quad (4)
\end{aligned}$$

where $v_p$, $v_c$, and $v_K$ are unit-free line-integrals of their corresponding basis coefficients, $\Phi(E) = [\phi_p(E), \phi_c(E), \phi_K(E)]$ and $\mathbf{v} = [v_p, v_c, v_K]^T$.

It is noteworthy mentioning that the basis functions for Photoelectric effect and Compton scattering in Eqs. (3) and (4) can be replaced by the linear attenuation coefficients of two basis materials such as water and bone. Also, the line-integrals can be interpreted as the equivalent thicknesses of the two basis materials. In this study, water, bone and 0.2% gadolinium were used as basis materials and Eq. (4) becomes

$$\begin{aligned}
\int \mu_a(\mathbf{r}, E) d\mathbf{r} &= v_w \mu_w(E) + v_b \mu_b(E) + v_g \mu_g(E) \\
&= \mathbf{U}(E)\mathbf{v}, \quad (5)
\end{aligned}$$

where $\mu_w(E)$, $\mu_b(E)$ and $\mu_g(E)$ are linear attenuation coefficients of water, bone and 0.2% gadolinium, respectively,

TABLE I
SIMULATION PARAMETERS FOR THE FOUR DETECTOR MODELS.

| Symbols | Meanings | Settings | Note |
| --- | --- | --- | --- |
| $S_0(E)$ | Probability mass function of the incident x-ray spectrum. | Null | FIG. 8 |
| $S_t(E)$ | Probability mass function of the transmitted x-ray spectrum. | Null | FIG. 9 |
| $a$ | Incident count rate | $5 \times 10^6$ Mcps/mm$^2$ | Null |
| $\Delta$ | Detection time frame | 20 ms | Null |
| $\tau$ | Detector deadtime | 30 ns | Null |
| $t_1, t_2$ | Parameters of the unipolar pulse shape | $0.5\tau, \tau$ | FIG. 1 and 10 |
| $t_1, t_2, t_3, b$ | Parameters of the bipolar pulse shape | $0.284\tau, 0.850\tau, 8.253\tau, -0.0635$ | FIG. 1 and 10 |
| $\delta_e$ | Gaussian standard deviation of the energy blurring | 2.6 keV | FIG. 5 |

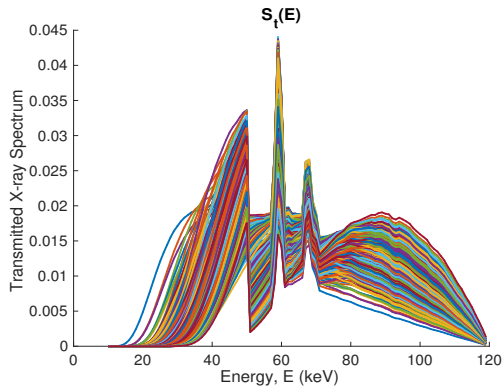

Fig. 9. Transmitted x-ray spectrum $S_t(E)$. Notice that each spectrum is normalized for better visualization.

$v_w$ [cm], $v_b$ [cm], and $v_g$ [cm] are the corresponding effective thicknesses, $\mathbf{U}(E) = [\mu_w(E), \mu_b(E), \mu_g(E)]$ and $\mathbf{v} = [v_w, v_b, v_g]^T$. In the simulation, we set $v_w \in [0, 12]$ cm with a step of 2 cm, $v_b \in [0, 2]$ cm with a step of 0.4 cm, and $v_g \in [0, 10]$ cm with a step of 2 cm. In total, we have 251 different attenuation paths. The transmitted x-ray spectrum $S_t$ is defined as follows:

$$S_t(E) = S_0(E) \times \exp(-\mathbf{U}(E)\mathbf{v}). \quad (6)$$

which is illustrated in FIG. 9.

*C. Step 3: Incident photons*

Let us denote the incident rate and detection time frame (also referred as reading) as $a$ and $\Delta$ respectively. In our study, $a$ and $\Delta$ were set to $5 \times 10^6$ counts per mm$^2$ per second (5 Mcps/mm$^2$) and $2 \times 10^{-2}$ seconds (20 ms), respectively. The size of pixel was assumed as 1 mm$^2$. This yields average $10^5$ photons per reading before attenuation. Then, the mean number of incident photons after attenuation $N_t$ is calculated as follows:

$$N_t = \sum S_t(E) \times a \times \Delta. \quad (7)$$

The counts in each trial were generated according to the Poisson distribution with mean equal to $N_t$, implemented with a MatLab function "poissrnd". The energy of each photon was sampled according to $S_t(E)$. The intervals between two subsequent photons was determined according to the exponential distribution with mean equal to $1/a_t$ by applying MatLab function "exprnd", while $a_t = a \sum S_t(E)$ is transmitted incident rate.

We simulated 10,000 measurements for each attenuation path and the incident photons in each trial carried the information of photon energy and incident time, and were sent the the pulse shaper described next.

*D. Step 4: Pulse shaper and detected signal*

We tested both the unipolar and bipolar pulse shapers (see FIG. 1). For the former, we simply implemented it as a isosceles triangle whose height scaled with photon energy and width was set to deadtime $\tau$ [21], which means that the peaking time $t_1 = 0.5\tau$ and pulse width $t_2 = \tau$. For the latter, we implemented it with a positive and a negative triangle. The height of the pulse was also proportional to photon energy, while the temporal extent remains the same. The positive lobe was an asymmetric triangle, reaching a maximum height at time $t_1 = 0.284\tau$ and dropping back to 0 at time $t_2 = 0.850\tau$. The negative tail was made a right triangle of a height $-0.0635$ of the positive maximum at time $t_2$ and ending at time $t_3 = 8.253\tau$ [12].

The deadtime $\tau$ in our simulations was $3^{-8}$ seconds (or 30 ns). Every incident photon produced a pulse according to either the unipolar or bipolar pulse shaper. The final signal was the aggregation of all pulses from all photons in each trial. We sample the signal at every $0.01\tau$. A fragment of the generated signal is illustrated in FIG. 10.

*E. Step 5: Signal detection*

The nonparalyzable and paralyzable detection modes were implemented in reference to [21]. According to previous parameter setting and Eq. (2), the deadtime loss ratio in the nonparalyzable and paralyzable cases were $[0.11\%, 13.04\%]$ and $[0.11\%, 13.93\%]$, respectively.

The counter of nonparalyzable detectors was repeatedly increased every deadtime by one count as long as the pulse height was above the baseline. The recorded energy was the maximum signal energy in each inactive state. No additional photons were registered while the detector was inactive. On the other hand, the paralyzable detector was realized by leaving the detector in an inactive state as long as the pulse height is above the baseline. The baseline energy in this study was set to 0 keV, although other values like 20 or 35 keV may be preferred since the existence of electronic noise/dark current.

In our simulations, the number of output counts at every 1 keV was recorded in the range [1, 240] keV, and any recorded event with energy larger than 240 keV was registered as 240

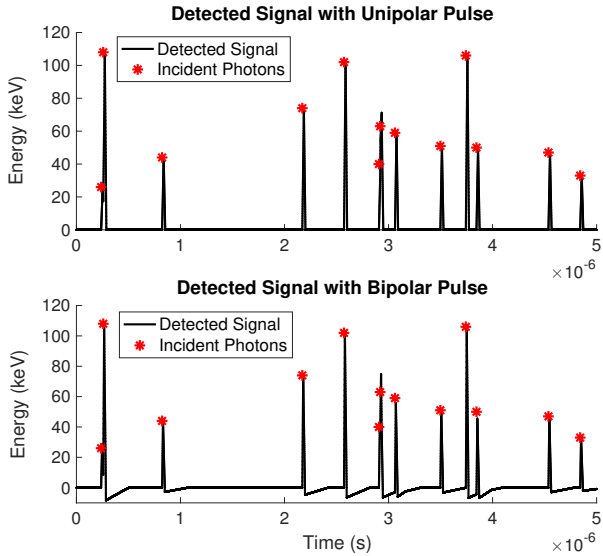

Fig. 10. Fragment of detected signal with unipolar and bipolar pulses respectively.

keV. Hence, we could re-bin the data in any fashion we like, without any new simulation run.

After we recorded the output counts, a zero mean, normally distributed value with a standard deviation $\delta_e$ of 2.6 keV is added to blur the output. In this sense, the energy resolution was 2.6 keV [12], [33].

The input data were feed to the neural network for better estimation of real data, utilizing data extracted from both the spectral and the trigger thresholds, which were [20, 39, 62, 81] keV and [120, 160] keV, respectively. For training, the target outputs were the corresponding ideal PCD data (with $\tau = 0$ and perfect energy resolution). The thresholds placement was not optimized in this initial investigation.

With the Monte Carlo simulation settings described above, we generated 251 different attenuation paths or transmittances, yielding 2,520,000 measurements in total, including the data with no attenuation for each of the four detector models.

## V. NUMERICAL RESULTS

First, we sorted the 252 attenuators (including the case of no attenuator) in terms of their DLRs. Then, we selected every 10 attenuators to produce test datasets and used the rest of the attenuators to synthesize training and validating datasets. In this way, the test data can uniformly span the whole space of PCD data. There were 250,000 test data points (about 10% of the total amount of simulated data). The ratio between training and validate data was 9 over 1. Furthermore, 5-fold cross validation was used to avoid overfitting with random initialization. In each validation cycle, the training and validate data were randomly selected. The mean performance metrics of 5 trained neural networks were obtained.

In the training phase, the Adam optimization method [34] was used to train the FCC artificial neural networks, the learning rate $lr$ is set to $1 \times 10^3$ with two exponential decay rates $\beta_1 = 0.9$ and $\beta_2 = 0.999$ for the moment estimates. The batch of 131,072 data points was used per iteration. The networks were trained with up to 1,000 epochs. In order to decrease the training time and avoid the overfitting, the training phase was terminated if the validation loss stopped decreasing in two consequent epochs. Finally, the dropout technique [35] was applied to improve the generalization of the networks.

Since the range of data is wide to diversified attenuation paths, the traditional loss function such as mean squared error (MSE) is not suitable. Hence, the mean absolute percentage error (MAPE) seems a good choice but it suffers from the zero denominator problem because of the zero count in low energy bins especially in the cases of high attenuation. Hence, we implemented a new loss function in percentage sense as:

$$Loss(y_{true}, y_{pred}) = \frac{1}{N} \sum_{i=1}^{N} \frac{\sum_{b=1}^{N_b} |y_{true,i}^b - y_{pred,i}^b|}{\sum_{b=1}^{N_b} y_{true,i}^b}, \quad (8)$$

where $N_b$ is the number of bins and $N$ is the batch size.

All the networks were implemented in Keras [36] with Tensorflow [37] as backend, and trained with a NVIDIA Titan Xp GPU.

### A. Different Network Structures

In this pilot study, the number of hidden neurons in each layer was made the same and set to 16, 32, 64, 128, and 256 respectively. As far as the number of layers is concerned, we started from one and increased one layer at a time until the validation loss stopped decreasing in two consequent simulations. The dropout rates were set to 0, 0.1, 0.3, 0.5, 0.7, and 0.9 respectively. Based on our results, dropout rate of 0.7 produced the overall best performance and was used as the default.

The validate loss measures of different networks are shown in FIG. 11. It can be observed in FIG. 11 that the loss performance was not very sensitive to the number of neurons or layers. For instance, either a two/three-layer FCC with 128 neurons or a two-layer FCC with 256 neurons in each layer would guarantee an acceptable validate loss in all cases, as summarized in TABLE II. Also, the validate loss for a paralyzable detector tended to be larger than that for a nonparalyzable model. It agrees with the fact that the number of counts recorded with a paralyzble detector are generally less because of the longer deadtime. In other words, the spectral distortion by paralyzable detectors would be greater, with the other conditions being identical. Finally, we focused on the FCC structures with the least validate loss (bold faced in TABLE II) in the subsequent simulations tests.

Next, the results with the selected FCC structures in the above settings were obtained from both training and test datasets. In order to measure the deviation of mean counts in each energy bin, we defined the relative error of count averages (RECA) as follows:

$$\text{RECA} = \frac{\sum_{b=1}^{N_b} |\frac{1}{N} \sum_{i=1}^{N} n_{i,org}^b - \frac{1}{N} \sum_{i=1}^{N} n_{i,new}^b|}{\sum_{b=1}^{N_b} (\frac{1}{N} \sum_{i=1}^{N} n_{i,org}^b)}, \quad (9)$$

where $N_b = 4$ is the number of energy bins, $N = 10,000$ is the number of trials for each attenuation path, $n_{i,org}^b$ is the true PCD counts without any PPE (deadtime $\tau = 0$) and $n_{i,new}^b$ is the distorted measurements with PPE or the recovered measurements from FCC. Note that the errors in terms of mean counts were measured over all trials.

TABLE II
VALIDATE LOSS MEASURES OF DIFFERENT FCC STRUCTURES

|  | NO. of Neurons | NO. of Layers | Validate Loss |
|---|---|---|---|
| UNP | 128 | 2 | 0.7619 |
|  | 128 | 3 | 0.7479 |
|  | **256** | **2** | **0.7063** |
| UP | 128 | 2 | 0.8005 |
|  | 128 | 3 | 0.7566 |
|  | **256** | **2** | **0.7535** |
| BNP | **64** | **6** | **0.7116** |
|  | 128 | 2 | 0.7714 |
|  | 128 | 3 | 0.7549 |
|  | 256 | 2 | 0.7386 |
| BP | 128 | 2 | 0.7866 |
|  | **128** | **3** | **0.7638** |
|  | 256 | 2 | 0.7741 |

TABLE III
RECA MEANS FOR DISTORTED AND FCC-CORRECTED DATA FOR DLR VALUES 0.1 %, 1 %, 10 % AND OVER TRAINING AND TEST DATA RESPECTIVELY. PCD AND FFC DENOTE DISTORTED AND FCC-CORRECTED DATA RESPECTIVELY.

|  | Detector | 0.1 % | 1 % | 10 % | Training | Test |
|---|---|---|---|---|---|---|
| PCD | UNP | 3.6386 | 4.1813 | 5.7984 | 3.6604 | 3.6604 |
|  | UP | 3.5803 | 4.4283 | 12.2036 | 4.1694 | 4.1694 |
|  | BNP | 3.7145 | 4.7271 | 8.3307 | 4.5074 | 4.5074 |
|  | BP | 3.6572 | 4.8383 | 11.0947 | 4.6888 | 4.6888 |
| FCC | UNP | 0.4018 | 0.2002 | 1.0060 | 0.3268 | 0.3268 |
|  | UP | 0.3297 | 0.2810 | 0.8879 | 0.3320 | 0.3320 |
|  | BNP | 0.3279 | 0.1406 | 1.2146 | 0.2835 | 0.2835 |
|  | BP | 0.4427 | 0.1625 | 0.9599 | 0.3481 | 0.3481 |

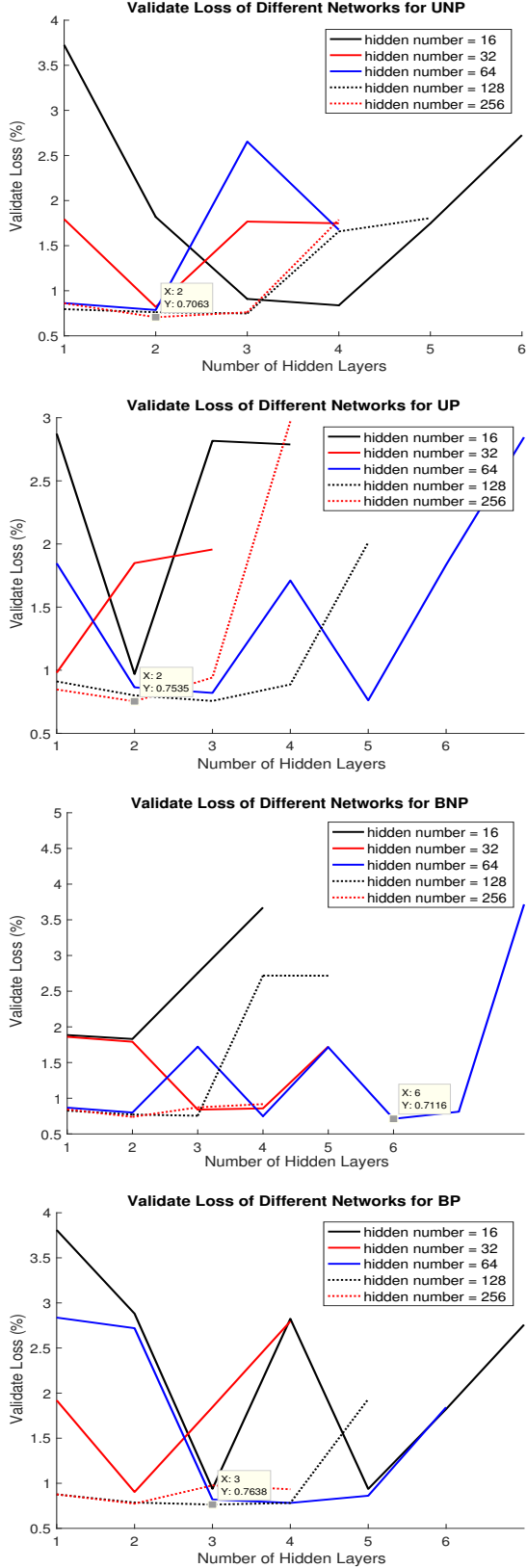

Fig. 11. Validate loss measures for different FCC structures.

The RECA comparisons of distorted and recovered data are shown in Fig. 12. As demonstrated in TABLE III, the RECA of corrected data was about one order of magnitude less than that of distorted data. Also, compared with raw PCD measurements, the RECA of corrected data was small when DLR was large, especially when DLR was larger than 1%. Last but not the least, the results from test datasets were in good agreement with that from training dataset, suggesting a great potential of the FCC network for generalization.

Actually, the mean RECA of training and test data is the same as shown in TABLE III. It means the trained model can be reliably used to correct new data different from training data without any significant performance loss.

It is clear that the FCC neural networks can recover the mean of distorted data. Then, it is natural to ask how much the variance of the corrected data is? Since the ideal counts of PCDs follows the Poisson distribution, the variance should be equal to the mean. Hence, we use the deviation of variance-to-mean ratio (DVMR) to evaluate the FCC model. The DVMR measure is defined as follows:

$$\begin{aligned} \text{DVMR} &= |1 - \text{VMR}| \quad (10) \\ &= |1 - \frac{\frac{1}{N}\sum_{i=1}^{N}(n_{i,new} - \frac{1}{N}\sum_{i=1}^{N}n_{i,new})^2}{\frac{1}{N}\sum_{i=1}^{N}n_{i,new}}|, \end{aligned}$$

while $n_{i,org} = \sum_{b=1}^{N_b} n_{i,org}^b$. One may suggest the variance as the metric. However, the variance varies dramatically with different attenuation paths. Hence, DVMR is more suitable to justify the performance. A perfect Poisson distribution will have VMR equal to 1 and DVMR equal to 0. Since the corrected data is still supposed to follow the Poisson distribution, neither larger nor smaller VMR/variance is preferable. Hence, the absolute difference between VMR and 1 (perfect Poisson case) was chosen to be our metric.

The performance of DVMR is illustrated in Fig. 13. Again, the DVMR of test data aligns with that of training data. Furthermore, it is observed that the DVMR of FCC-corrected data was slightly smaller for nonparalyzable detectors than for paralyzable detectors. If the correction process was simply scaling the distorted data by multiplying a constant $k$ to achieve desired mean, the variance would be scaled by a factor of $k^2$. However, the performance of the FCC model suggests that the recovery process is beyond linear transformation. Additionally, the real detectors behave in the way between the nonparalyzable and paralyzable modes, it is estimated that the FCC model in that case should give a similar DVMR performance.

Next, we will validate the effect of our proposed loss function (8). Given the space limitation, in the following we will only show the results for the UNP detector as an example under the same FCC setting specified in TABLE II unless otherwise mentioned.

### B. Different Loss Functions

The proposed loss function was compared with two other loss functions often used in neural networks: mean square error (MSE) and mean absolute percentage error (MAPE). The performance of different loss functions is illustrated in FIG. 14.

Generally speaking, the DVMR measures of the FCC model trained with the three loss functions were comparable. However, the proposed loss function gave the least RECA. For MSE, it is observed that the FCC model yielded a much better performance for high-DLR data. The reason is that the MSE weighted more on high-count data. On the other hand, MAPE calculated the percentage error in each energy bin, and suffered from the divide-by-zero problem. In contrast, our proposed loss function addressed these problems by calculating the fraction between the difference and the number of true counts over all the energy bins.

### C. Different Numbers of Trigger Thresholds

The number of trigger thresholds was evaluated for its impact on the data correction. The number of trigger thresholds was set to 0, 1, 2, and 3 respectively. In the last three cases, the energy thresholds were 120, 120 and 160, as well as 120, 160, and 200 respectively. The results are summarized in FIG. 15 and TABLE IV. Again, there was no significant difference among these settings in term of DVMR. On the other hand, it can be seen that although giving better performance for low-DLR data, in the no-trigger-threshold case the largest RECA was produced for DLR being close to 10%. To be specific, compared with the default setting, the RECA in the no-trigger-threshold case was increased by about 35% over the entire dataset. It proves the utility of the proposed trigger threshold method in the recovery process. Furthermore, the performance with different trigger thresholds did not differ substantially. The reason may be due to the relative small distortion in the dataset (maximum DLR was about 10%). Hence, there were relatively few counts above 160 or 200 keV, and the effects of more trigger thresholds are limited. The performance of proposed methodology on heavier distorted data is worth to investigated in the further.

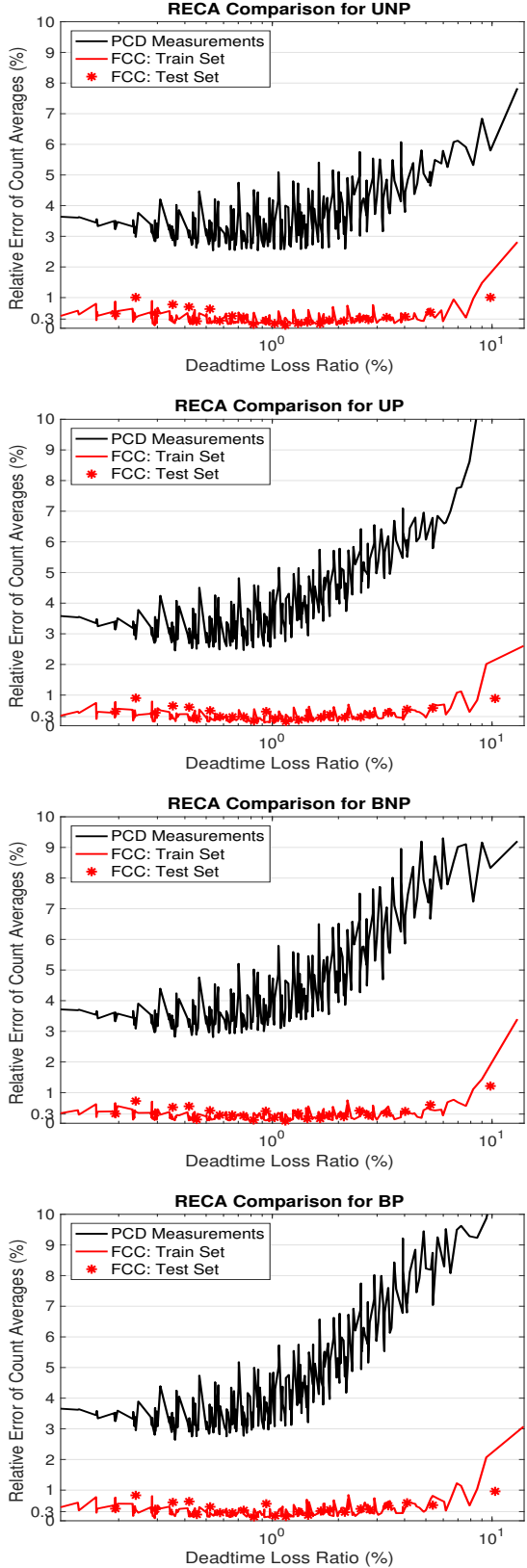

Fig. 12. RECA means for distorted and FCC-corrected data respectively.

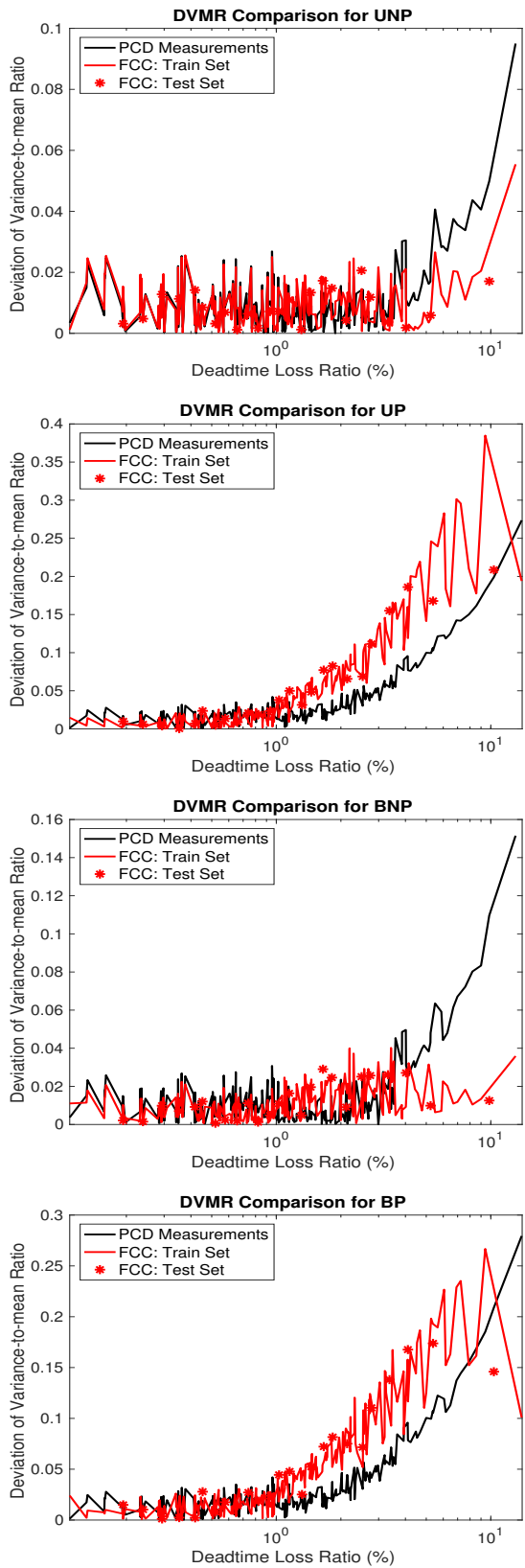

Fig. 13. DVMR means for distorted and FCC-corrected data respectively.

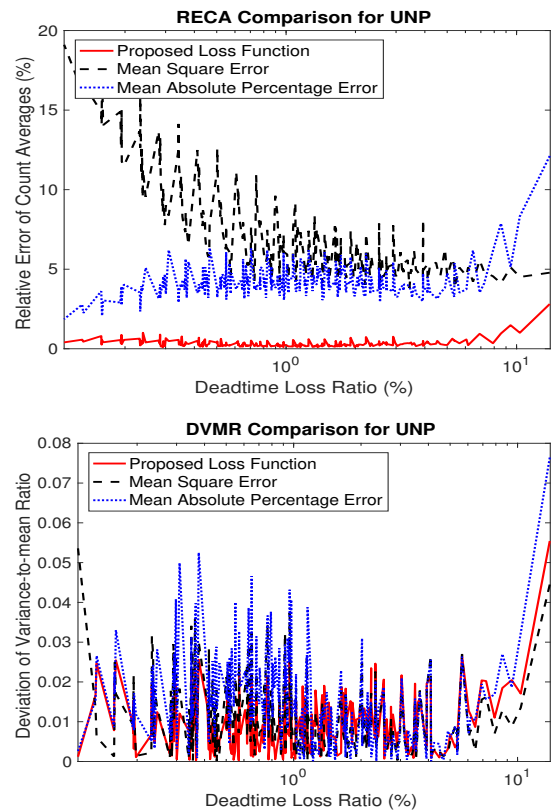

Fig. 14. Comparison for different loss functions.

TABLE IV
RECA MEANS FOR DIFFERENT NUMBER OF TRIGGER THRESHOLDS WITH UNP DETECTOR WITH DLR VALUE 0.1 %, 1 % AND 10 %, AND OVER TRAINING AND TEST DATASETS RESPECTIVELY.

| NO. of Trigger Thresholds | 0 | 1 | 2 | 3 |
|---|---|---|---|---|
| DLR = 0.1% | 0.3512 | 0.4807 | 0.4018 | 0.3207 |
| DLR = 0.1% | 0.2064 | 0.2518 | 0.2002 | 0.2297 |
| DLR = 0.1% | 2.5154 | 1.4000 | 1.4801 | 1.4164 |
| Training | 0.4410 | 0.3156 | 0.3268 | 0.3234 |
| Test | 0.4410 | 0.3156 | 0.3268 | 0.3234 |

### D. Different Energy Resolutions

As mentioned in Section IV, the non-perfect energy resolution would blur the recorded spectrum. Hence, we compared the behaviors when the data inputs are from detectors with perfect and 2.6 keV resolution. The performance is illustrated in FIG. 16. It is seen that the RECA of PCD data with non-perfect resolution is obviously larger than that with perfect resolution. One the other hand, there was no clear difference in FCC-corrected data with the different energy resolutions. It means that our methodology corrected the data distortion from not only the PPE but also from the non-perfect energy resolution.

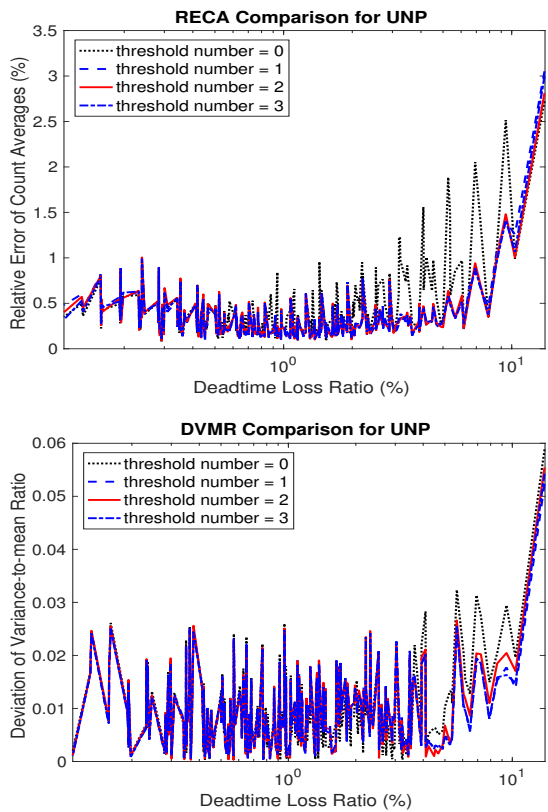

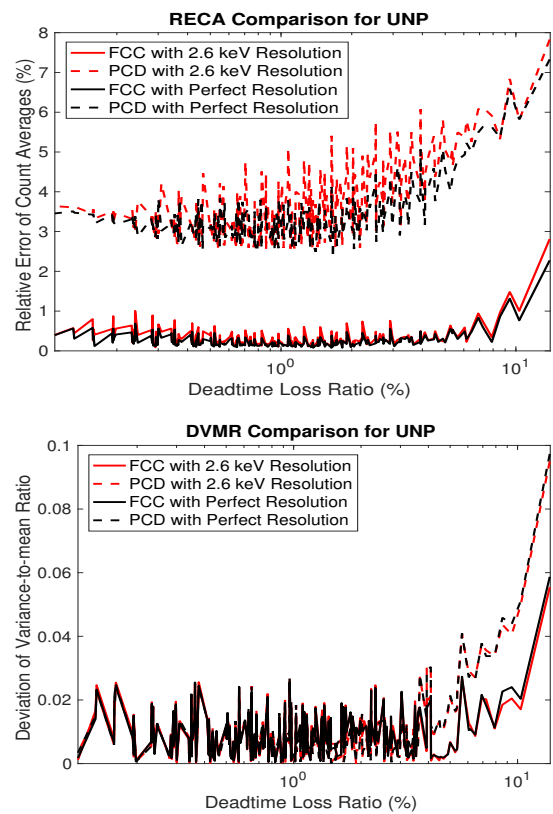

Fig. 15. Comparison for different numbers of trigger thresholds.

Fig. 16. Comparison for different energy resolutions.

### E. Performance with Different Training Set Sizes

Finally, we tested the FCC model trained with different training dataset sizes. We sampled the test dataset by every 10, 4, 3 and 2 attenuators respectively. Then, we sampled the training dataset by every 3, 4, and 10 attenuators respectively. The performance is illustrated in FIG. 17. It is clear that the FCC model worked well until the training set percentage was around 70%, and the FCC model collapsed when the ratio between training and test dataset size reached 50%. These data serves as a guideline about the dataset size we should collect with real photon counting detectors.

## VI. DISCUSSIONS AND CONCLUSION

In this article, we have proposed a FCC ANN based approach improve the distorted PCD measurements by suppressing the pulse pileup effect. The correction performance has been evaluated with data simulated via Monte Carlo simulation.

Although the results on simulated data was promising, the performance can be further improved by optimizing the network structure and the training strategy. For example, we assumed the number of hidden neurons are the same for each layer. However, a better structure may be found by comparing networks of different numbers of layers and neurons. On the other hand, the training parameters can be further refined, including the batch size, the learning rate and the decay setting.

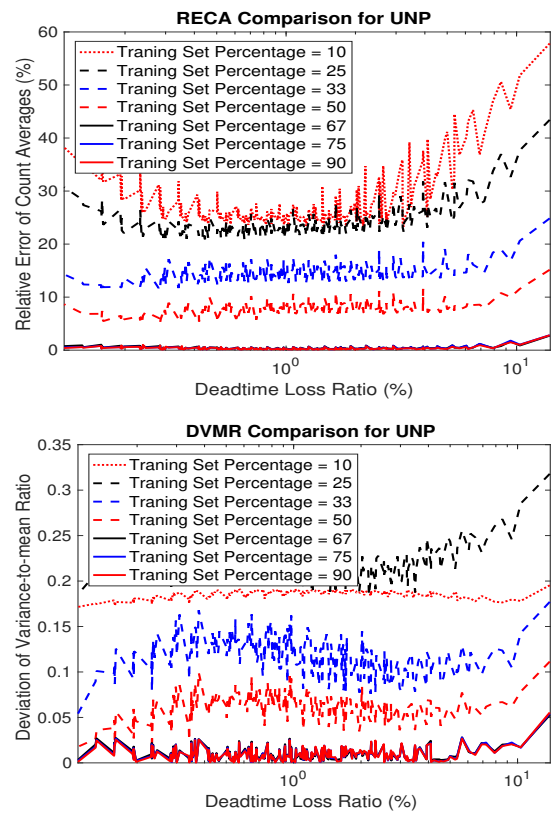

Fig. 17. Comparison for different training dataset sizes.

The current method only corrects the data from one pixel at a time. Although the process is fast, we intend to adapt our method from the pixel/1-D model to the image/2-D model in order to fully utilize data correlation and the parallel processing ability of GPUs. In fact, the adaption can be done by implementing a shallow version of DenseNet [26] with a small convolution kernel without any pooling layer.

In stead of simulated data with simplified detector models, the validation can be performed with real PCD data, which is critical before applying our proposed method in clinical and preclinical applications. One might think that the collection of training data $(\mathbf{X}, \mathbf{Y})$, especially the original undistorted data $\mathbf{Y}$, is difficult or even unrealistic in real PCDs experiments. Accurately, it is fair easy by following the proposed steps described as below.

Step 1: Measure the mean and standard deviation of PPE-free data at a low x-ray flux
The goal of the low flux is to collect data with no/rare pulse pileup. It could be achieved by decreasing the tube current, adding filtration, or increasing the distance from the x-ray tube to attenuators. In order to make sure the measurements are free from the pileup, we could decrease the count rate until there is no counts with energy larger than the tube energy (e.g. 120 keV). Then, the raw data $(\mathbf{Y}_{low})$ is recorded, and the mean $(mean(\mathbf{Y}_{low}))$ as well as standard deviation $(std(\mathbf{Y}_{low}))$ are calculated separately for each energy bin.

Step 2: Estimate the mean and standard deviation of PPE-free data at the desired x-ray flux
The mean $(mean(\mathbf{Y}_{des}))$ and standard deviation $(std(\mathbf{Y}_{des}))$ could be estimated by scaling $mean(\mathbf{Y}_{low})$ and $std(\mathbf{Y}_{low})$ with respect to the current-to-flux proportionality constant (e.g. $k$ in [22]), the attenuation coefficient of filtration, or the geometry disciplines.

Step 3: Collect PPE-compromised data at the desired x-ray flux
The distorted data $(\mathbf{X})$ is directly collected with the desired experimental setting. Note that the data should contain the counts from both spectral and trigger thresholds.

Step 4: Estimate PPE-free data at the desired x-ray flux
First of all, the data $\mathbf{X}$ is re-binned to $\mathbf{X}_{des}$ with only spectral energy bins. Then, a normalization algorithm, which takes $\mathbf{X_{des}}$, $mean(\mathbf{Y}_{des})$ and $std(\mathbf{Y}_{des})$ as inputs and generates PPE-free data $\mathbf{Y}_{des}$ as outputs, is applied. The algorithm is summarized in Algorithm 1.

After the 4-step process, we could have ($\mathbf{X}$ and its estimated undistorted counterpart $\mathbf{Y}_{des}$). Then, the data pair is used to train the neural networks by assuming $\mathbf{Y} = \mathbf{Y}_{des}$. As an example, we test the accuracy of estimated $\mathbf{Y}_{des}$ using the previous simulated data with DLR equal to 0.1%, 1%, 10% respectively. The deviation of the $j$th measurement between $\mathbf{Y}$ and $\mathbf{Y}_{des}$ is evaluated by $100 \times ||\mathbf{Y}_j - \mathbf{Y}_{des,j}||/||\mathbf{Y}_j||$. The results are summarized in TABLE V. It can be observed that the maximum deviation is around 5% while the mean deviation is smaller than 1.5% in all cases. The results demonstrate the proposed data collection process is accurate enough to construct dataset for the neural network training.

TABLE V
DEVIATION BETWEEN $\mathbf{Y}$ AND $\mathbf{Y}_{des}$

| Detector | DLR (%) | 0.1 % | 1 % | 10 % |
|---|---|---|---|---|
| UNP | Min Deviation (%) | 0.0755 | 0.0418 | 0.0211 |
|  | Max Deviation (%) | 5.4215 | 2.0533 | 0.8467 |
|  | Mean Deviation (%) | 1.4623 | 0.6099 | 0.2522 |
| UP | Min Deviation (%) | 0.1230 | 0.0348 | 0.0211 |
|  | Max Deviation (%) | 5.5246 | 1.9704 | 0.9958 |
|  | Mean Deviation (%) | 1.4664 | 0.6164 | 0.2868 |
| BNP | Min Deviation (%) | 0.1049 | 0.0189 | 0.0227 |
|  | Max Deviation (%) | 5.3821 | 1.9892 | 0.9932 |
|  | Mean Deviation (%) | 1.4690 | 0.6242 | 0.2732 |
| BP | Min Deviation (%) | 0.0855 | 0.0422 | 0.0321 |
|  | Max Deviation (%) | 5.4619 | 2.1667 | 1.1698 |
|  | Mean Deviation (%) | 1.4714 | 0.6253 | 0.2914 |

---
**Algorithm 1** Data Normalization
---
**Input:** $\mathbf{X}_{des}$, $mean(\mathbf{Y}_{des})$, $std(\mathbf{Y}_{des})$
**Output:** $\mathbf{Y}_{des}$
    **for** $i = 1$ to $N$ **do**     ▷ $N$ is the number of energy bins
        $\mathbf{y} \leftarrow \mathbf{X}_{des}^i$     ▷ $\mathbf{X}_{des}^i$ is measurements in energy bin $i$
        **if** $std(\mathbf{y}) > 0$ **then**
            $\mathbf{y} \leftarrow \mathbf{y}/std(\mathbf{y}) \times std(\mathbf{Y}_{des}^i)$ ▷ Normalize the std of $\mathbf{y}$ to std of $\mathbf{Y}_{des}$ in $i$th energy bin
        **end if**
        $\mathbf{y} \leftarrow \mathbf{y} - mean(\mathbf{y}) + mean(\mathbf{Y}_{des}^i)$     ▷ Normalize the mean of $\mathbf{y}$ to mean of $\mathbf{Y}_{des}$ in $i$th energy bin
        $\mathbf{Y}_{des}^i \leftarrow \mathbf{y}$ ▷ Assign $\mathbf{y}$ to the measurements of $\mathbf{Y}_{des}$ in $i$th energy bin
    **end for**
---

In this study, we have only considered the pulse pileup effect and assumed data independence for each pixel and every energy bin. However, there exist other degradation factors (such as charge sharing [38] and K-escape x-rays) and strong data correlation across pixels and energy bins. For example, when charge sharing is involved, the resultant electrons will be divided by adjacent pixels with energies lower than the original one. Recently, Taguchi et al. released a software tool "Photon Counting Toolkit (PcTK)" [33], [39]. PcTK takes key degradation factors and correlation between PCD pixels into consideration and can be used to synthesize noisy PCD projections. With this toolkit, we plan to extend our approach for charge sharing correction.

In conclusion, we have proposed a fully-connected cascade artificial neural network to improve the PCD measurements distorted due to the pulse pileup effect. In stead of complicated mathematical modeling, our proposed model learns the recovery process from training data. The simulation results has convincingly demonstrated that the trained networks can correct data very well. Specifically, our proposed approach offers similar improvements for different detector models, and has great potentials for preclinical and clinical applications.

We are working to optimize and generalize this neural network approach as a new PCD data processing strategy.

ACKNOWLEDGMENT

The authors would like to thank NVIDIA Corporation for the donation of Titan Xp GPU used for this research.